\documentclass{ieeeaccess}
\usepackage{pgfplots}

\usepackage{cite}
\usepackage{amsmath,amssymb,amsfonts}
\usepackage{algorithmic}
\usepackage{graphicx}
\usepackage{textcomp}
\usepackage{float}
\usepackage{amsmath}
\usepackage{amssymb}
\usepackage{booktabs}
\usepackage{makecell}
\usepackage{bm}

\usepackage{listings}
\usepackage{xcolor}
\usepackage{hyperref}

\usepackage{tikz}
\usepackage{cuted}  
\usepackage{capt-of}
\usepackage[acronym, nonumberlist, nogroupskip]{glossaries}

\makeglossaries

\newacronym{CLI}{CLI}{Command Line Interface}
\newacronym{DQC}{DQC}{Distributed Quantum Computing}
\newacronym{EPR}{EPR}{Einstein-Podolsky-Rosen}
\newacronym{GHZ}{GHZ}{Greenberger-Horne-Zeilinger}
\newacronym{HPC}{HPC}{High-Performance Computing}
\newacronym{ISA}{ISA}{Instruction Set Architecture}
\newacronym{LOC}{LOC}{lines of code}
\newacronym{MPI}{MPI}{Message Passing Interface}
\newacronym{NISQ}{NISQ}{Noisy Intermediate-Scale Quantum}
\newacronym{OS}{OS}{Operating System}
\newacronym{QKD}{QKD}{Quantum Key Distribution}
\newacronym{QMPI}{QMPI}{Quantum Message Passing Interface}
\newacronym{QPU}{QPU}{Quantum Processing Unit}
\newacronym{SDK}{SDK}{Software Development Kit}
\newacronym{SPMD}{SPMD}{Single Program Multiple Data}

\pgfplotsset{compat=1.18}

\usetikzlibrary{shapes.symbols, positioning, shadows, calc, arrows.meta, fit, backgrounds}

\NewSpotColorSpace{PANTONE}
\AddSpotColor{PANTONE} {PANTONE3015C} {PANTONE\SpotSpace 3015\SpotSpace C} {1 0.3 0 0.2}
\SetPageColorSpace{PANTONE}%

\definecolor{codegreen}{rgb}{0,0.6,0}
\definecolor{codegray}{rgb}{0.5,0.5,0.5}
\definecolor{codepurple}{rgb}{0.58,0,0.82}
\definecolor{codeblue}{rgb}{0.0, 0.2, 0.6}
\definecolor{backcolour}{rgb}{0.96,0.96,0.96}

\lstdefinestyle{mystyle}{
    backgroundcolor=\color{backcolour},   
    commentstyle=\color{codegreen},
    keywordstyle=\color{codeblue}\bfseries,
    numberstyle=\tiny\color{codegray},
    stringstyle=\color{codepurple},
    basicstyle=\ttfamily\scriptsize, 
    breakatwhitespace=false,         
    breaklines=true,                 
    captionpos=b,                    
    keepspaces=true,                 
    numbers=left,                    
    numbersep=5pt,                  
    showspaces=false,                
    showstringspaces=false,
    showtabs=false,                  
    tabsize=2,
    frame=single, 
    rulecolor=\color{black!20}
}

\lstdefinelanguage{NetQASM}{
    keywords={set, qalloc, init, h, cnot, meas, qfree, store, load, create_epr, wait_all, ret, bne, beq, br},
    keywordstyle=\color{blue}\bfseries,
    keywords=[2]{R0, R1, R2, @0, @1, @2, @3}, 
    keywordstyle=[2]\color{orange!80!black},
    sensitive=false, 
    comment=[l]{\#}, 
    morestring=[b]", 
}

\makeatletter
\AtBeginDocument{\DeclareMathVersion{bold}
\SetSymbolFont{operators}{bold}{T1}{times}{b}{n}
\SetSymbolFont{NewLetters}{bold}{T1}{times}{b}{it}
\SetMathAlphabet{\mathrm}{bold}{T1}{times}{b}{n}
\SetMathAlphabet{\mathit}{bold}{T1}{times}{b}{it}
\SetMathAlphabet{\mathbf}{bold}{T1}{times}{b}{n}
\SetMathAlphabet{\mathtt}{bold}{OT1}{pcr}{b}{n}
\SetSymbolFont{symbols}{bold}{OMS}{cmsy}{b}{n}
\renewcommand\boldmath{\@nomath\boldmath\mathversion{bold}}}
\makeatother

\def\BibTeX{{\rm B\kern-.05em{\sc i\kern-.025em b}\kern-.08em
    T\kern-.1667em\lower.7ex\hbox{E}\kern-.125emX}}

\begin{document}
\history{Date of publication xxxx 00, 0000, date of current version xxxx 00, 0000.}
\doi{XX.ZZZZ/ACCESS.YYYY.XXXXXXX}
\title{NetQMPI: An MPI-Inspired library for programming Distributed Quantum Applications over Quantum Networks using NetQASM SDK}
\author{\uppercase{F. Javier Cardama}\authorrefmark{1} \IEEEmembership{Student Member, IEEE}, and
\uppercase{Tomás F. Pena}\authorrefmark{1,2} \IEEEmembership{Senior Member, IEEE}}

\address[1]{Centro Singular de Investigación en Tecnoloxías Intelixentes (CiTIUS), Universidade de Santiago de Compostela, Santiago de Compostela, 15782, Spain}
\address[2]{Departamento de Electrónica e Computación, Universidade de Santiago de Compostela, Santiago de Compostela, 15782, Spain}
\tfootnote{This work was funded by the European Union EuroHPC program with grant agreement 101194491 and by MICIU/AEI/10.13039/501100011033 with project number PCI2025-163229, the Agencia Estatal de Investigación (Spain) (PID2022-141623NB-I00), the Consellería de Cultura, Educación e Ordenación Universitaria Galician Research Center accreditation 2024–2027 ED431G-2023/04, and the European Regional Development Fund (ERDF).}

\markboth
{F. Javier Cardama \headeretal: NetQMPI: An MPI-Inspired library for programming Distributed Quantum Applications ...}
{F. Javier Cardama \headeretal: NetQMPI: An MPI-Inspired library for programming Distributed Quantum Applications ...}

\corresp{Corresponding author: F. Javier Cardama (email: javier.cardama@usc.es).}

\begin{abstract}
Distributed Quantum Computing (DQC) is essential for scaling quantum algorithms beyond the limitations of monolithic NISQ devices. However, the current software ecosystem forces developers to manually orchestrate low-level network resources, such as entanglement generation (EPR pairs) and classical synchronization, leading to verbose, error-prone, and non-scalable code. 
This paper introduces \textbf{NetQMPI}, a high-level Python framework that adapts the Message Passing Interface (MPI) standard to the quantum domain using the Single Program Multiple Data (SPMD) paradigm. Built as a middleware over the NetQASM SDK, NetQMPI abstracts the underlying physical topology, automating network initialization and resource management through a unified \texttt{Communicator} interface. 
We propose semantic point-to-point primitives and novel collective operations—such as \texttt{expose} and \texttt{unexpose}—that address the constraints of the No-Cloning Theorem by leveraging multipartite entanglement for data distribution. 
Our comparative analysis demonstrates that NetQMPI decouples algorithmic logic from network size, reducing the code complexity for generating an $N$-node GHZ state from $\mathcal{O}(N^2)$ to constant complexity $\mathcal{O}(1)$. Furthermore, the framework ensures backend agnosticism, enabling the seamless execution of high-level applications on rigorous physical simulators, such as NetSquid (via SquidASM), and future quantum hardware adhering to the NetQASM standard.
\end{abstract}


\titlepgskip=-21pt

\maketitle

\section{Introduction}
\label{sec:introduction}

In recent years, quantum computing has emerged as a computational paradigm capable of offering exponential speedups for problems intractable on classical machines~\cite{Buhrman1998QuantumComputation, Markov2014LimitsComputation}, such as integer factorization~\cite{Shor1999Polynomial-timeComputer} and unstructured search~\cite{Grover1996ASearch}. However, realizing these algorithms at scale requires a substantial number of qubits and robust error correction capabilities, which remain challenging for current hardware in the \gls*{NISQ} era~\cite{Preskill2018QuantumBeyond}.

To overcome these scaling limitations, \gls{DQC} has emerged as a promising path. By interconnecting multiple \glspl{QPU} via a quantum network, it is possible to create a virtual cluster capable of executing large-scale algorithms~\cite{Barral2025ReviewComputing, Caleffi2024DistributedSurvey, VanMeter2008ArchitectureAlgorithm}. However, orchestrating computations across such a distributed infrastructure introduces engineering complexities~\cite{Ovide2023MappingArchitectures}. Unlike classical networks, quantum networks require the management of fragile resources such as entanglement, \gls{EPR} pairs, and the tight synchronization of classical and quantum control planes~\cite{Wehner2018QuantumAhead, Illiano2022QuantumSurvey, Azuma2023QuantumInternet}.

Nevertheless, the transition from monolithic to distributed quantum algorithms introduces significant software engineering challenges~\cite{Loke2022FromOverview}. Most existing software tools require researchers to program at the physical or link layer, where they must explicitly manage elementary operations~\cite{Cardama2026NetQIR:Computing, Diaz-Camacho2025BenchmarkingEmulators,Barral2025ReviewComputing}. At this level, developers are often burdened with defining hardware-specific parameters such as photon emission timing, fiber attenuation, and memory decoherence , while also manually orchestrating network protocols like entanglement generation (\gls{EPR} pairs), routing, and the tight synchronization of classical and quantum control planes~\cite{Cacciapuoti2020QuantumComputing}.

This low-level approach not only increases development complexity but also couples the algorithmic logic to specific hardware topologies or simulation backends, severely limiting code portability and maintainability. In the realm of classical \gls*{HPC}, this challenge was successfully addressed by the \gls*{MPI} standard, which unified distributed programming, typically following the \gls*{SPMD} paradigm. While theoretical proposals for a \gls*{QMPI} exist~\cite{Haner2021DistributedQmpi}, the ecosystem still lacks a functional, standardized implementation that bridges the gap between high-level algorithmic design and the execution on rigorous quantum network stacks.

\subsection{Contributions}
In this work, we present \textbf{NetQMPI}, a high-level MPI-inspired Python library for programming distributed quantum applications over quantum networks. NetQMPI acts as a middleware layer over the standard NetQASM \gls{SDK}, allowing developers to write distributed quantum algorithms using high-level semantics while ensuring execution on rigorous network simulators.

The main objectives and novel contributions of this work are:

\begin{itemize}
    \item \textbf{Adoption of the SPMD Paradigm:} We propose a shift from role-based scripting to a unified Single Program Multiple Data model. By injecting logical \texttt{rank} identifiers, NetQMPI enables a single source file to orchestrate complex protocols across $N$ nodes, thereby decoupling code complexity from network size.
    
    \item \textbf{Abstraction of Physical Resources:} We introduce the \texttt{QMPICommunicator} and semantic primitives (e.g., \texttt{qsend}, \texttt{qrecv}) that abstract the management of EPR sockets and classical control planes, which allows researchers to focus on algorithmic logic rather than the intricate details of entanglement generation and fidelity management.
    
    \item \textbf{Novel Collective Operations:} We define and implement quantum-specific collective operations, such as \texttt{expose} and \texttt{unexpose}. These primitives address the constraints of the No-Cloning Theorem by leveraging multipartite entanglement to share quantum information across the communicator without physical copying.
    
    \item \textbf{Backend-Agnostic Execution:} By targeting the standardized NetQASM \gls*{ISA}, NetQMPI serves as a unified interface that can execute applications on high-fidelity simulators (such as NetSquid via SquidASM) and is forward-compatible with future quantum hardware, bridging the gap between abstract algorithm design and realistic physical simulation.
\end{itemize}

This paper is organized as follows. Section~\ref{sec:background} introduces the related work including tools for DQC programming and simulation. Section~\ref{subsec:netqasm} provides an overview of the NetQASM framework, distinguishing between its \gls{ISA} and the \gls{SDK}, while highlighting the programming complexity inherent in the current ecosystem. Section~\ref{sec:netqmpi} details the architecture of NetQMPI as well as the creation process on NetQASM SDK and the high-level semantic functions provided. Section~\ref{sec:comparative} presents a comparative analysis of the various tools in the literature described in the related work section, and, finally, Section~\ref{sec:conclusions} concludes the paper.
\section{Related Work}
\label{sec:background}
Recent advances in quantum algorithms have driven considerable efforts in developing software tools for programming and simulating distributed quantum systems. This section presents the most relevant tools in this domain, distinguishing between high-level abstractions and low-level physical simulators.

\subsection{Software Tools for Programming Distributed Quantum Systems}
\label{subsec:software-tools}

To manage the complexity of distributed quantum states, several frameworks have been proposed to abstract the underlying physical layer.

\textbf{QuNetSim} \cite{Diadamo2021QuNetSim:Networks} provides a high-level Python API for simulating network protocols (e.g., \gls*{QKD}, teleportation) without dealing with hardware physics.
\begin{itemize}
    \item \textit{Strengths:} It offers an extremely gentle learning curve and built-in implementations of standard protocols.
    \item \textit{Weaknesses:} It is purely a simulator with no path to physical hardware execution; its abstraction level hides the resource management details (like entanglement fidelity) required for realistic system programming.
\end{itemize}

\textbf{Interlinq} \cite{Parekh2021QuantumComputing} offers a high-level abstraction layer built upon the QuNetSim communication framework, designed to automate circuit partitioning and message scheduling across distributed nodes. 
\begin{itemize}
    \item \textit{Strengths:} It simplifies distributed execution by automating circuit partitioning and teleportation scheduling, allowing monolithic circuits to be run across multiple nodes (``circuit knitting'', i.e., splitting a quantum program to fit onto smaller devices linked by classical communication~\cite{Piveteau2024CircuitCommunication}) without manual intervention for small-scale systems.
    \item \textit{Weaknesses:} It suffers from severe scalability issues, where execution time and memory usage grow exponentially with the number of qubits and nodes. Additionally, its automatic partitioner lacks advanced optimization techniques, leading to redundant teleportations and communication overhead, while imposing rigid constraints on node configurations.
\end{itemize}

\textbf{QMPI}, proposed by Häner et al.~\cite{Haner2021DistributedQmpi}, introduces the theoretical foundation for applying the Message Passing Interface (MPI) standard to quantum computing.
\begin{itemize}
    \item \textit{Strengths:} It leverages the mature MPI standard (commands like \texttt{Send}, \texttt{Recv}, \texttt{Bcast}) familiar to the HPC community.
    \item \textit{Weaknesses:} There is no known implementation or functional application of this proposal, as the work is limited to defining the semantics. Furthermore, it does not address how to execute collective operations such as \texttt{Bcast} or \texttt{Allgather}; these primitives inherently imply data copying in classical MPI, which conflicts with the No-Cloning Theorem, and the proposal offers no mechanism to resolve this.
\end{itemize}

\subsection{Quantum Network Simulators}
\label{subsec:network-simulators}
This category encompasses tools designed to simulate any quantum network. These tools allow the execution of platform-independent code, often relying on lower-level engines for physical accuracy.

\textbf{SimulaQron}~\cite{Dahlberg2018SimulaQronaSoftware} focuses on the distributed nature of the network. It runs a simulation where each network node is a separate process on the host machine, communicating via classical sockets to imitate a real distributed architecture.
\begin{itemize}
    \item \textit{Strengths:} It enforces a strict separation of memory between nodes, making it ideal for verifying client-server logic and distributed application flows.
    \item \textit{Integration:} While it acts as a standalone distributed backend, it is part of the QuTech ecosystem (a suite of standardized quantum networking tools developed at TU Delft) and was a precursor to modern execution environments.
\end{itemize}

\textbf{SquidASM}~\cite{Dahlberg2022NetQASMaInternet} is the specialized execution tool for NetQASM applications. It acts as a bridge between the application layer and the physical layer.
\begin{itemize}
    \item \textit{Strengths:} It allows developers to run NetQASM routines on a simulated network with high fidelity.
    \item \textit{Integration:} SquidASM is explicitly built on top of NetSquid (see next subsection). It translates the high-level instructions from the SDK into discrete events that NetSquid can process, providing a realistic noise model while maintaining a programmable interface.
\end{itemize}

\subsection{Discrete-Event Simulators}
\label{subsec:discrete-event}
At the lowest level of abstraction, these tools model the physical transmission of qubits, time-dependent noise, and hardware components (memories, repeaters, fibers).

\textbf{NetSquid}~\cite{Coopmans2021NetSquidEvents} is the state-of-the-art platform for modeling quantum networks at the physical layer.
\begin{itemize}
    \item \textit{Strengths:} It handles precise timing of photon emission, fiber loss, and quantum memory decoherence. It serves as the physics engine powering higher-level tools like SquidASM.
    \item \textit{Weaknesses:} The complexity of defining hardware components makes it cumbersome for general application programming; it is designed for network architects rather than algorithm developers.
\end{itemize}

\textbf{SeQUeNCe} \cite{Wu2021SeQUeNCe:Networks} is a customizable discrete-event simulator with a strong focus on the network layer (routing, resource management).
\begin{itemize}
    \item \textit{Strengths:} Excellent for studying network topology and protocol stacks, and available as open-source.
    \item \textit{Weaknesses:} Like NetSquid, it presents a high barrier to entry for software developers simply wanting to write quantum applications using standard programming paradigms.
\end{itemize}

\section{NetQASM SDK Overview}
\label{subsec:netqasm}

The foundation of the work presented in this paper is the NetQASM framework~\cite{Dahlberg2022NetQASMaInternet}, which can be distinguished into two main components: the Instruction Set Architecture (ISA) and the Software Development Kit (SDK).

\subsection{NetQASM ISA}
NetQASM defines a low-level, platform-independent \gls{ISA} explicitly designed for quantum networks. It serves as a common language that abstracts the heterogeneity of underlying hardware. Programs compiled into NetQASM assembly can be executed on various backends, ranging from high-fidelity physical simulators like SquidASM (based on NetSquid) or SimulaQron. This architecture separates the quantum logic from the control hardware, allowing for hybrid quantum-classical execution.

To illustrate the granularity of control required by this architecture, Figure \ref{fig:netqasm-isa} presents a snippet of Alice's operations. The execution flow begins with explicit resource management, where the program manually assigns a value to a classical register (\texttt{set R0 0}) to identify, allocate (\texttt{qalloc}), and initialize (\texttt{init}) a local control qubit. Subsequently, the entanglement generation process requires preparing the arguments in specific memory addresses—storing the number of pairs at \texttt{@0} and the type at \texttt{@1}—before triggering the \texttt{create\_epr} instruction. Notably, the code must include a \texttt{wait\_all} instruction to explicitly block execution until the physical hardware confirms the entanglement generation. Finally, the script manually maps the remote entangled qubit to a register (\texttt{set R1 1}), performs a local two-qubit gate (\texttt{cnot}), measures the result into a classical register (\texttt{meas}), and explicitly frees the quantum memory with \texttt{qfree}.

\begin{figure}[tbp]
\centering
\begin{lstlisting}[language=NetQASM]
# NetQASM Assembly (Alice Side)
set R0 0         # Register for Qubit ID
qalloc R0        # Allocate Control Qubit
init R0          # Initialize to |0>

# -- Entanglement Generation --
store 1 @0       # Store arg: number of pairs
store 0 @1       # Store arg: type
create_epr 0 1 0 0 @0 
wait_all @0      # Explicit wait for hardware

set R1 1         # Manually assign ID to EPR qubit
cnot R0 R1       # Local CNOT
meas R1 R2       # Measure EPR into Register R2
qfree R1         # Free EPR qubit resource
\end{lstlisting}
\caption{Snippet of NetQASM ISA Assembly corresponding to Alice's operations. It highlights the low-level management of registers (e.g., \texttt{R0}, \texttt{R1}) and memory addresses (e.g., \texttt{@0}) required by the architecture.}
\label{fig:netqasm-isa}
\end{figure}

\begin{figure*}[t]
    \centering
    \begin{minipage}{0.49\linewidth}
        \textbf{Alice (Control Node)}
        \begin{lstlisting}[language=Python, xleftmargin=0em]
# Create a socket to send classical information
socket = Socket("Bob", "Alice", log_config=log_config)

# Create an EPR socket for entanglement generation
epr_socket = EPRSocket("Bob")

with NetQASMConnection(epr_sockets=[epr_socket]) as conn:
    # Create a qubit to teleport
    q = Qubit(conn)

    # Create EPR pairs
    epr = epr_socket.create_keep()[0]

    # Teleport
    q.cnot(epr)
    q.H()
    m1 = q.measure()
    m2 = epr.measure()

# Send the correction information
m1, m2 = int(m1), int(m2)

socket.send_structured(StructuredMessage("Corrections", (m1, m2)))
    \end{lstlisting}
    \end{minipage}
    \hfill 
    \begin{minipage}{0.49\linewidth}
        \textbf{Bob (Target Node)}
        \begin{lstlisting}[language=Python, numbers=none, xleftmargin=0pt]
# Create a classical socket
socket = Socket("Alice", "Bob")

# Create an EPR socket
epr_socket = EPRSocket("Alice")

with NetQASMConnection(epr_sockets=[epr_socket]) as conn:
    # 1. Receive EPR qubit
    epr = epr_socket.recv_keep()[0]

    # 2. Flush to sync
    conn.flush()

    # 3. Get and apply the corrections
    m1, m2 = socket.recv_structured().payload

    if m2 == 1:
        epr.X()

    if m1 == 1:
        epr.Z()

    # 4. Final flush to sync
    conn.flush()
    \end{lstlisting}
    \end{minipage}
    \captionof{figure}{Implementation of a teleport protocol using NetQASM SDK. The developer is responsible for manually orchestrating the EPR socket creation, classical message passing, and synchronization (flush).}
    \label{fig:netqasm-cnot}
\end{figure*}

\subsection{NetQASM SDK}
To facilitate programming without writing raw assembly, the NetQASM SDK provides a Python interface for writing applications for specific network roles (e.g., ``Alice'' and ``Bob''). A central abstraction in the SDK is the \textit{EPR Socket}.
By opening an EPR socket between two nodes, the programmer can request the generation of entanglement (e.g., creating Bell pairs) utilizing the underlying network link layer.

Figure~\ref{fig:netqasm-cnot} demonstrates the implementation of a teleport protocol using the NetQASM SDK, highlighting the steps required to transfer a qubit state from Alice to Bob.

\subsection{Programming Complexity}
While the SDK abstracts the physical generation of entanglement, it does not provide high-level primitives for qubit transmission.
Crucially, to transfer a quantum state from one node to another, the generation of an EPR pair is merely the first step. The programmer must manually implement the full quantum protocol communication \textit{ad hoc}.

For example, to send a single qubit via teleportation~\cite{Bennett1993TeleportingChannels} using the SDK, the developer must explicitly:
\begin{enumerate}
    \item Create or request entanglement via the EPR Socket and wait for completion.
    \begin{itemize}
        \item Alice: line 12.
        \item Bob: line 9.
    \end{itemize}
    \item Perform local gates (CNOT, Hadamard) between the payload qubit and the entangled qubit.
    \begin{itemize}
        \item Alice: lines 15--16.
    \end{itemize}
    \item Measure the qubits and obtain the classical outcomes.
    \begin{itemize}
        \item Alice: Lines 17--18.
    \end{itemize}
    \item Send the classical bits to the receiver using a separate classical channel.
    \begin{itemize}
        \item Alice: Line 23.
    \end{itemize}
    \item Receiver side: read these bits and apply the corresponding Pauli corrections ($X$ or $Z$) to recover the state.
    \begin{itemize}
        \item (Bob: Line 15--21).
    \end{itemize}
\end{enumerate}

This requirement to manually orchestrate classical and quantum synchronization for every transfer leads to verbose and error-prone code. This limitation is the primary motivation for NetQMPI, which encapsulates these complex workflows into simple, atomic communication primitives.
\section{NetQMPI}
\label{sec:netqmpi}

This section introduces NetQMPI, a high-level Python library designed to simplify the development of distributed quantum applications. Inspired by the standardized Message Passing Interface (MPI) used in classical HPC, NetQMPI adopts the Single Program Multiple Data (SPMD) paradigm, which enables developers to write a single, platform-independent source code that orchestrates quantum operations across multiple network nodes without manually managing the underlying physical entanglement resources.

\subsection{Software stack}
\label{subsec:stack}

\begin{figure}[htbp]
    \centering
    \includegraphics[width=0.8\linewidth]{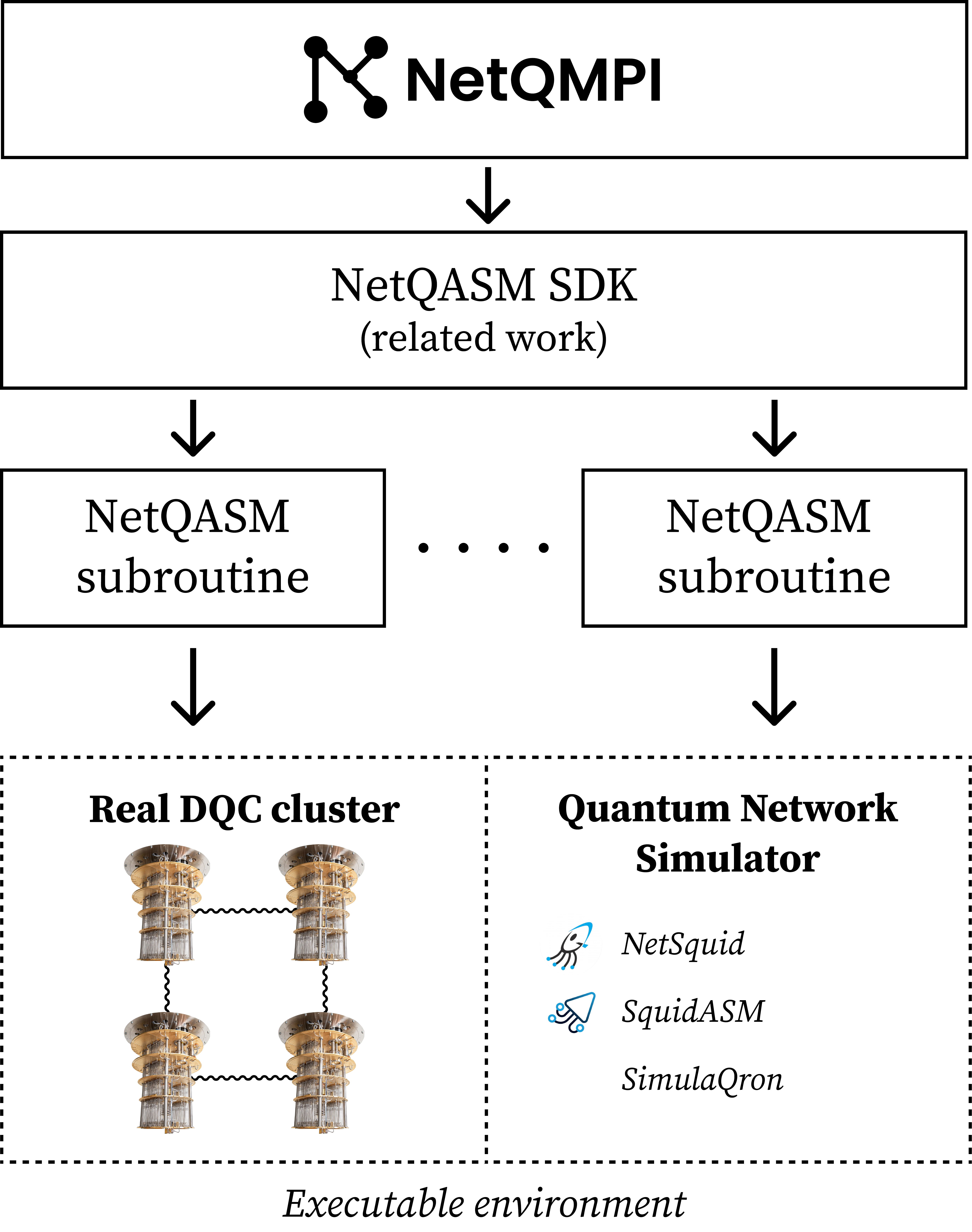}
    \caption{The NetQMPI software stack. The library wraps the verbose NetQASM SDK primitives, providing a unified API for the application layer while relying on the standard NetQASM ISA to interface with backend simulators, such as SquidASM or SimulaQron, or real DQC clusters supporting NetQASM.}
    \label{fig:netqmpi_stack}
\end{figure}

Figure~\ref{fig:netqmpi_stack} illustrates the architectural position of NetQMPI within the quantum network software ecosystem. NetQMPI acts as a middleware layer that sits directly on top of the NetQASM SDK.

A primary objective of this architecture is to bridge the gap between high-level algorithmic design and platform-specific execution. By abstracting the NetQASM SDK, NetQMPI hides the complexity of connection management and explicit instruction compilation. The user application interacts solely with NetQMPI primitives (e.g., \texttt{qsend}, \texttt{qrecv}), which are then translated at runtime into the corresponding NetQASM subroutines—handling \texttt{EPRSocket} lifecycle, \texttt{NetQASMConnection} management, and classical control flow transparently.

Because these primitives compile down to the standardized NetQASM ISA, NetQMPI is inherently backend-agnostic. This abstraction allows developers to write code that is completely decoupled from the underlying infrastructure, enabling the execution of the same source code on both \textit{simulated networks} and \textit{physical quantum networks} without modification.

Furthermore, this design leverages the robust capabilities of the ecosystem. Specifically, it allows users to utilize high-fidelity simulators like NetSquid (via the SquidASM interface) without facing its steep learning curve. While NetSquid is the state-of-the-art for modeling physical noise and network delays, programming it directly for application development is notoriously complex. NetQMPI enables researchers to validate algorithms using NetSquid's realistic physics engine while maintaining a high-level, abstract programming model.

\subsection{Construction of NetQMPI from NetQASM SDK}
\label{subsec:construction}

NetQMPI\footnote{The complete source code is available at: \url{https://github.com/netqir/netqmpi}} is constructed as a wrapper that fundamentally alters the workflow of the NetQASM SDK. To achieve the transition from the role-based scripts required by the SDK to the Single Program Multiple Data (SPMD) paradigm, NetQMPI relies on a specific internal architecture. Figure~\ref{fig:netqmpi_tree} illustrates the organization of the source code.

The core orchestration is handled by the \texttt{external.py} module, which manages the multiprocess environment mapping. The network state and automated entanglement generation are encapsulated within the \texttt{communicator} package. Finally, the specific communication logic is organized under \texttt{primitives}, divided into \texttt{p2p} and \texttt{collective} modules, whose specific implementations will be detailed in Subsections \ref{subsec:p2p} and \ref{subsec:collective}, respectively.

\begin{figure}[tbp]
\centering
\begin{lstlisting}[basicstyle=\ttfamily\scriptsize, frame=single, xleftmargin=2em]
netqmpi/
|-- sdk/
|   |-- external.py
|   |-- communicator/
|   |   |-- communicator.py
|   |-- primitives/
|   |   |-- collective/
|   |   |-- p2p/
\end{lstlisting}
\caption{Directory structure of the NetQMPI library. The \texttt{external.py} module orchestrates the multiprocess execution, while \texttt{communicator.py} manages the automated entanglement generation.}
\label{fig:netqmpi_tree}
\end{figure}

\subsubsection{Single Program Architecture}
In the standard NetQASM SDK, developers must write $N$ separate source files (e.g., \texttt{app\_alice.py} and \texttt{app\_bob.py}), one for each node in the network. NetQMPI unifies this into a single file, following the SPMD paradigm.

To enable this, NetQMPI utilizes the abstractions of \textit{rank} and \textit{size}. For readers unfamiliar with HPC terminology, the \textit{rank} is a unique integer identifier assigned to each node in the distributed network, ranging from $0$ to $size-1$, while the \textit{size} represents the total number of participating nodes in the communicator. These values are injected automatically by the library into the user's code at runtime, allowing the developer to define the logic for all nodes in a single function signature.

The user script must accept \texttt{rank} and \texttt{size} as arguments, which are then used to initialize the \texttt{QMPICommunicator}, as shown in Figure~\ref{fig:rank_size_injection}.

\begin{figure}[tbp]
\begin{lstlisting}[language=Python]
from netqmpi.sdk.communicator import QMPICommunicator

def main(app_config=None, rank=0, size=1):
    # The rank is injected by NetQMPI
    # and used to initialize the local communicator
    COMM_WORLD = QMPICommunicator(rank, size, app_config)
    print(f"Hello, rank={rank} of {size} processes")
\end{lstlisting}
\caption{An example of the rank and size variable injection with the global communicator creation.}
\label{fig:rank_size_injection}
\end{figure}

\begin{figure}[btp]
    \centering
    \resizebox{\linewidth}{!}{\begin{tikzpicture}[
    node distance=1cm and 0.5cm,
    auto,
    block/.style={
        rectangle, 
        draw=blue!50, 
        fill=blue!10, 
        text width=2cm, 
        text centered, 
        rounded corners, 
        minimum height=1.5cm,
        font=\scriptsize,
        label={[anchor=north east, xshift=-1pt, yshift=-1pt, text=blue!80]north east:$\star$}
    },
    sdk/.style={
        rectangle, 
        draw=orange!80, 
        fill=orange!20, 
        text width=2.5cm, 
        text centered, 
        rounded corners, 
        minimum height=1.2cm,
        font=\footnotesize\bfseries,
        label={[anchor=north east, xshift=-1pt, yshift=-1pt, text=orange!90!black]north east:$\blacktriangle$}
    },
    process/.style={
        rectangle, 
        draw=green!50, 
        fill=green!10, 
        text width=1cm, 
        text centered, 
        minimum height=1cm,
        font=\scriptsize,
        label={[anchor=north east, xshift=1pt, yshift=1pt, text=green!60!black]north east:$\bullet$}
    },
    terminal/.style={
        rectangle,
        draw=black!50,
        fill=black!5,
        text width=2.6cm,
        font=\ttfamily\scriptsize,
        align=left,
        minimum height=1cm,
        label={[anchor=north east, xshift=-2pt, yshift=-2pt, text=black!60, font=\tiny]north east:\texttt{>\_}}
    },
    wrapper/.style={
        rectangle, 
        draw, 
        dashed, 
        fill=gray!5,
        inner sep=0.2cm, 
        rounded corners
    },
    line/.style={draw, -Latex, thick},
    legend_text/.style={
        font=\scriptsize, 
        anchor=west
    }
]

    
    \node [terminal] (user) {\textbf{\$} netqmpi -n 3 f.py};
    \node [block, right=of user] (cli) {\textbf{cli.py} \\ (Parse args)};
    \node [block, right=of cli] (external) {\textbf{external.py} \\ (Define roles, inject Rank/Size)};
    
    \node [sdk, below=1.5cm of external] (sdk) {NetQASM SDK \\ (Process Spawner)};
    
    \node [process, below=1cm of sdk] (proc2) {Rank 1 \\ (Bob)};
    \node [process, left=0.2cm of proc2] (proc1) {Rank 0 \\ (Alice)};
    \node [process, right=0.2cm of proc2] (proc3) {Rank 2 \\ (Charlie)};
    
    \path [line] (user) -- (cli);
    \path [line] (cli) -- (external);
    \path [line] (external.south) -- node[left, font=\scriptsize, align=left, xshift=-0.1cm] {NetQASM app config + \\ Modified script} (sdk.north);
    \path [line] (sdk.south) -- (proc1.north);
    \path [line] (sdk.south) -- (proc2.north);
    \path [line] (sdk.south) -- (proc3.north);
    
    \begin{pgfonlayer}{background}
        \node[wrapper, fit=(cli) (external), label={[font=\scriptsize]above:\textbf{NetQMPI Middleware Layer}}] (wrapperbox) {};
    \end{pgfonlayer}

    
    \matrix [
        draw=black!20, 
        fill=white, 
        rounded corners, 
        below=3.2cm of user.west, 
        anchor=west,              
        row sep=6pt,              
        column sep=5pt            
    ] (legend) {
        \node[fill=blue!10, draw=blue!50, minimum size=1em, label={center:$\star$}] {}; & 
        \node[legend_text] {NetQMPI Module}; \\
        \node[fill=orange!20, draw=orange!80, minimum size=1em, label={center:$\blacktriangle$}] {}; & 
        \node[legend_text] {External Dependencies}; \\ 
        \node[fill=green!10, draw=green!50, minimum size=1em, label={center:$\bullet$}] {}; & 
        \node[legend_text] {Runtime Process}; \\
    };
    
    \node [above=2pt of legend.north west, anchor=south west, font=\scriptsize\bfseries] {Legend:};

\end{tikzpicture}}
    \caption{Execution flow of a NetQMPI application. The \texttt{external.py} module orchestrates the multiprocess environment, injecting rank information into the unified user script and delegating the process management to the NetQASM SDK.}
    \label{fig:execution_flow_compact}
\end{figure}
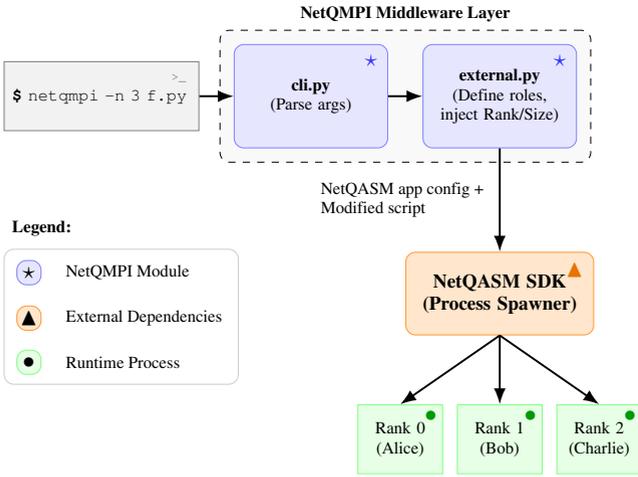

When the user executes this script using the Command Line Interface (CLI), the \texttt{cli.py} module parses the \texttt{-n} argument (number of processes) and triggers the \texttt{external.py} dispatcher. This dispatcher spawns the simulation processes using the NetQASM SDK, passing the specific rank to each instance. For example, executing the code with three nodes results in:

\begin{lstlisting}[language=bash, numbers=none, frame=single, backgroundcolor=\color{white}]
$ netqmpi -n 3 script.py
Hello, rank=0 of 3 processes
Hello, rank=2 of 3 processes
Hello, rank=1 of 3 processes
\end{lstlisting}

As illustrated in Figure~\ref{fig:execution_flow_compact}, the execution flow is orchestrated by \texttt{external.py}. It parses the unified script, defines the network roles based on the requested size, and injects the corresponding rank information into the application's entry point. Finally, it hands over the configuration to the underlying NetQASM SDK, which manages the physical spawning of the runtime processes (e.g., Alice, Bob, Charlie).

\subsubsection{Automated Network Initialization}
The core abstraction that manages the logical connectivity between the distributed nodes is the \texttt{QMPICommunicator} class, defined in the \texttt{communicator} package.

\begin{figure}[tbp]
    \centering
    \includegraphics[width=\linewidth]{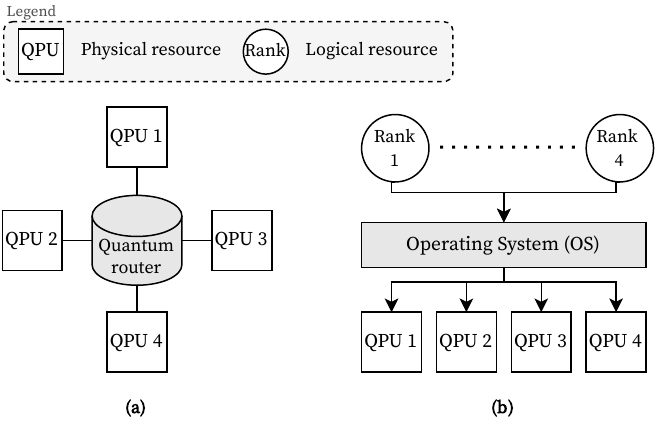}
    \caption{Logical abstraction for scheduling physical resources. (a) An example of physical resources interconnected in a physical quantum network. (b) The Communicator provides a logical layer where abstract Ranks are assigned to processes, and the OS/Middleware automatically handles the mapping to the available physical QPUs.}
    \label{fig:communicator_abstraction}
\end{figure}

Conceptually, a communicator acts as a virtualization layer that groups a set of logical processes connected through a logical network. This construct abstracts the underlying complexity of the infrastructure, where logical processes (identified by ranks) are mapped to physical resources (Quantum Processing Units) via a physical interconnection network. This relationship is visually described in Figure~\ref{fig:communicator_abstraction}. Figure~\ref{fig:communicator_abstraction}(a) depicts the physical reality where QPUs are connected via a Quantum Router, while Figure~\ref{fig:communicator_abstraction}(b) illustrates how the software stack (OS/Middleware) maps the high-level Logical Ranks ($1 \dots n$) to these physical resources transparently.

This approach mirrors the architecture of classical distributed computing, allowing developers to focus on the algorithmic logic rather than the physical topology. For a comprehensive definition of this abstraction model in distributed quantum computing, we refer the reader to the architectural framework presented in~\cite{Cardama2026NetQIR:Computing}, specifically Figure 6.

From an implementation perspective, the communicator automates resource management that is currently handled manually in the NetQASM SDK. In NetQASM, establishing a network requires the user to instantiate a \texttt{NetQASMConnection} and define specific \texttt{EPRSocket} objects for every pair of communicating nodes, leading to an $O(N^2)$ initialization complexity in fully connected topologies.
NetQMPI resolves this by performing the following automated steps during initialization:
\begin{enumerate}
    \item It assigns a unique \texttt{rank} to each logical process.
    \item It iterates through the network configuration and automatically instantiates the necessary mesh of \texttt{EPRSocket} between all participating ranks.
    \item It stores these sockets in an internal routing table within the communicator instance, making the management of entanglement resources transparent to the user.
\end{enumerate}

Furthermore, NetQMPI handles the synchronization of the quantum control flow. While the NetQASM SDK requires manual \texttt{flush()} invocations, NetQMPI primitives include implicit flushing mechanisms or offer explicit \texttt{flush()} operations to ensure state consistency across the network.

\subsection{Point to point functions}
\label{subsec:p2p}

The fundamental requirement of a distributed quantum system is the ability to transfer quantum states between nodes. NetQMPI encapsulates the standard Quantum Teleportation Protocol into two atomic primitives: \texttt{qsend} and \texttt{qrecv}.

Crucially, these primitives represent a shift towards \textbf{high-level semantic communication}. From the programmer's perspective, the operation is defined simply as sending a specific qubit to a destination rank. All references to low-level hardware resources—such as explicit EPR pair generation, entanglement fidelity, specific QPU identifiers, or classical control signaling—are completely abstracted away. The developer focuses solely on the logical flow of quantum information.

\subsubsection{The \texttt{qsend} Primitive}
The \texttt{qsend(qubit, dest\_rank)} function abstracts the complex role of the sender (``Alice''). When this semantic function is invoked, the library transparently handles the following underlying operations:
\begin{enumerate}
    \item Looks up the pre-established \texttt{EPRSocket} associated with the logical \texttt{dest\_rank} in the routing table.
    \item Requests the creation of an EPR pair from the network layer.
    \item Performs the local Bell measurement (CNOT and Hadamard gates) between the user's payload qubit and the entangled qubit.
    \item Measures the qubits and automatically transmits the classical measurement outcomes (2 bits) to the destination rank using the classical control plane.
\end{enumerate}

\subsubsection{The \texttt{qrecv} Primitive}
The \texttt{qrecv(source\_rank)} function abstracts the role of the receiver (``Bob''). It returns a handle to the received quantum state, allowing the program to continue execution immediately. Internally, the library:
\begin{enumerate}
    \item Waits for the hardware confirmation of the entanglement generation.
    \item Listens for the classical correction bits sent by \texttt{source\_rank}.
    \item Applies the necessary Pauli-X and Pauli-Z corrections to recover the original state fidelity.
\end{enumerate}

This encapsulation drastically simplifies the development process. As shown in Figure~\ref{fig:point-to-point}, the complex coordination required by the raw SDK (discussed in Section~\ref{subsec:netqasm}) is reduced to intuitive, single-line commands.

\begin{figure}[tb]
\begin{lstlisting}[language=Python]
def main(app_config=None, rank, size):
    # Initialize the communicator
    comm = QMPICommunicator(rank, size, app_config)

    if rank == 0:
        # Sender Logic (Alice)
        q = Qubit(comm.connection) 
        q.H() # Prepare state |+>
        
        # Semantic send: "Send qubit q to Rank 1"
        # No EPR management required
        comm.qsend(q, 1) 
        print("Rank 0: Qubit sent")

    elif rank == 1:
        # Receiver Logic (Bob)
        # Semantic receive: "Get qubit from Rank 0"
        q_received = comm.qrecv(0)
        
        # The qubit is ready for use
        m = q_received.measure()
        print(f"Rank 1: Qubit received, measure: {m}")
\end{lstlisting}
\caption{Example of point to point code using the \texttt{qsend} and \texttt{qrecv} primitives of NetQMPI.}
\label{fig:point-to-point}
\end{figure}

\subsection{Collective operations}
\label{subsec:collective}

A significant advantage of the MPI paradigm is the ability to perform collective operations involving multiple nodes simultaneously. While point-to-point communication allows for specific data transfer, collective primitives abstract common data movement patterns, reducing code verbosity and potential errors. NetQMPI implements these operations by leveraging the underlying point-to-point primitives.

\subsubsection{Scatter and Gather}
The \texttt{qscatter} and \texttt{qgather} primitives allow for the distribution and collection of quantum states among a group of processes.

The \texttt{qscatter} operation takes a list of qubits residing in a source process (\textit{root}) and distributes them sequentially to all other processes in the communicator. Conversely, \texttt{qgather} collects qubits from all processes and moves them to a destination register in the root process.
From an implementation standpoint, these operations are abstractions of a loop of point-to-point transfers. For instance, a \texttt{qscatter} is functionally equivalent to the root node executing a \texttt{qsend} inside a \texttt{for} loop targeting each rank in the communicator.

\begin{figure}[tbp]
    \centering
    \includegraphics[width=\linewidth]{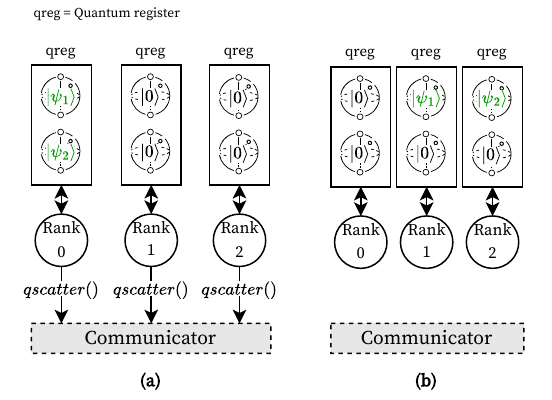}
    \caption{Visual representation of the \texttt{qscatter} operation. (a) Initial state where all qubits reside in the root process. (b) Distributed state where qubits have been moved to their respective ranks. The operations rely on iterative point-to-point teleportations.}
    \label{fig:scatter_gather}
\end{figure}

Figure~\ref{fig:scatter_gather} illustrates the movement of qubits during these operations. In Figure~\ref{fig:scatter_gather}(a), the initial state is shown where the root holds a register of qubits. Figure~\ref{fig:scatter_gather}(b) shows the result of a scatter operation (or the start of a gather), where each qubit has been teleported to its corresponding rank.

A practical implementation is shown in Figure~\ref{lst:scatter_example}. In this example, the root process initializes a list of qubits in the superposition state $|+\rangle$ (using Hadamard gates) and distributes one qubit to each node in the network using \texttt{qscatter}. Finally, each node measures its local qubit independently.

\begin{figure}
\begin{lstlisting}[language=Python]
def main(app_config=None, rank, size):
    comm = QMPICommunicator(rank, size, app_config)
    conn = comm.connection
    ROOT_RANK = 0
    qubits = []

    # 1. Root initializes the qubits
    if rank == ROOT_RANK:
        qubits = [Qubit(conn) for _ in range(size)]
        for q in qubits: 
            q.H() # Apply Hadamard
    
    # 2. Scatter: Distribute from Root to all
    # Returns the specific qubit assigned to this rank
    local_qs = comm.qscatter(qubits, ROOT_RANK)

    m = local_qs[0].measure())
\end{lstlisting}
\caption{Example of the \texttt{qscatter} collective primitive in NetQMPI. The root process initializes a list of qubits in the superposition state $|+\rangle^{\otimes n}$, scatters them to all ranks, and each rank measures its local qubit.}
\label{lst:scatter_example}
\end{figure}

\subsubsection{Expose and Unexpose}
In classical MPI, the \texttt{MPI\_Bcast} operation copies data from one process to all others. However, in quantum computing, the \textit{No-Cloning Theorem} prohibits creating independent copies of an arbitrary unknown quantum state. Therefore, a direct quantum broadcast is physically impossible.

To address this, NetQMPI introduces the \texttt{expose} and \texttt{unexpose} primitives, as proposed in~\cite{Cardama2026NetQIR:Computing}. Instead of copying the state, the \texttt{expose} operation creates a shared context where a specific qubit (or set of qubits) becomes conceptually accessible to other processes for collective operations, without violating the no-cloning principle.

Mathematically, the objective of \texttt{expose} is to transition the system from an initial state where the root process $R_0$ holds an arbitrary superposition
\begin{equation}\label{eq:initial_state_pre_expose}
    |\psi\rangle_{initial} = \alpha|0\rangle_{R_0} + \beta|1\rangle_{R_0},
\end{equation}
to a global state where this quantum information is distributed across all $n$ participating ranks ($R_0, \dots, R_n$).
\begin{equation}\label{eq:final_state_post_expose}
    |\psi\rangle_{exposed} = \alpha(|0\rangle_{R_0}\otimes\cdots\otimes|0\rangle_{R_n}) + \beta(|1\rangle_{R_0}\otimes\cdots\otimes |1\rangle_{R_n})
\end{equation}

\begin{figure}[tbp]
    \centering
    \includegraphics[width=\linewidth]{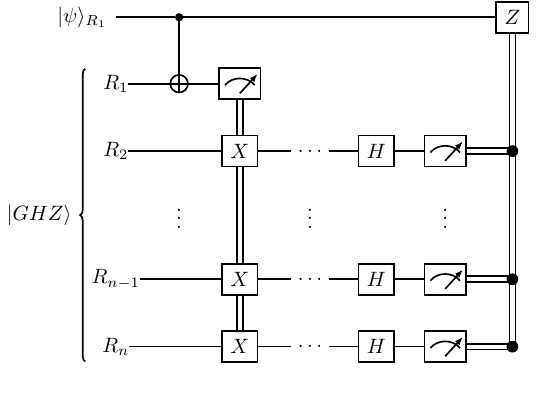}
    \caption{Quantum circuit representation of the \texttt{expose} and \texttt{unexpose} operation. The root qubit is entangled with a pre-shared GHZ state across all nodes, followed by local measurements and classical corrections to distribute the quantum information.}
    \label{fig:expose_circuit}
\end{figure}

This shared state enables collective controlled operations on the quantum information distributed across all nodes. The implementation relies on entangling the root qubit with a pre-shared GHZ state across all nodes, followed by local measurements and classical corrections to distribute the quantum information, as illustrated in Figure~\ref{fig:expose_circuit}.

Crucially, because this operation alters the accessibility scope of the qubits, it is mandatory to release the resource once the collective computation is finished. The \texttt{unexpose} primitive serves this purpose, returning the system to a consistent state where the qubits are once again private to their owner or properly measured.
\section{Comparative and Use Cases}
\label{sec:comparative}

To evaluate the effectiveness of the proposed abstractions, we perform a comparative analysis against the current state-of-the-art tools for distributed quantum programming. We selected the generation of a distributed Greenberger–Horne–Zeilinger (GHZ) state across $N$ nodes as the benchmark scenario. This task is chosen because it requires multi-partite entanglement and coordinated control flow, serving as a standard "Hello World" for the Quantum Internet.

\subsection{Benchmark Definition}
The goal is to generate the maximally entangled state $|\text{GHZ}\rangle = \frac{1}{\sqrt{2}}(|00\dots0\rangle + |11\dots1\rangle)$ shared among $N$ distinct network nodes, where each rank holds exactly one qubit.

\begin{figure}
    \centering
    \includegraphics[width=\linewidth]{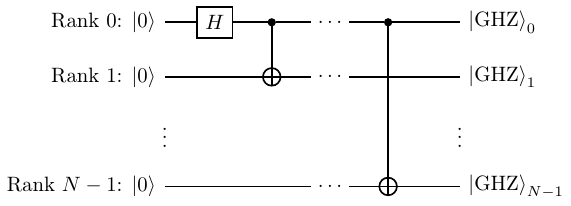}
    \caption{Distributed GHZ State Generation across $N$ nodes. The root node (Rank 0) prepares a qubit in superposition and performs distributed CNOTs to entangle the qubits at the target nodes (Ranks $1 \dots N-1$). Each Remote CNOT involves entanglement distribution and classical communication.}
    \label{fig:ghz_benchmark}
\end{figure}

As shown in Figure~\ref{fig:ghz_benchmark}, the protocol involves the following steps:
\begin{enumerate}
    \item One root node (e.g., Rank 0) prepares one qubit in superposition ($|+\rangle$).
    \item The root node performs a sequence of distributed CNOT operations targeting the qubits of the remaining $N-1$ nodes.
    \item In a distributed setting without shared memory, every ``Remote CNOT'' implies a complex teleportation protocol or entanglement swapping routine involving classical communication and corrections.
\end{enumerate}

\subsection{Qualitative Analysis: Programming Complexity}

We compare the implementation workflow in NetQMPI against three categories of tools: High-level Simulators (QuNetSim), Raw SDKs (NetQASM/SquidASM), and Discrete-Event Engines (NetSquid/SeQUeNCe).

\textbf{1. NetQMPI (This Work):}
Following the SPMD paradigm, the solution is written in a \textit{single source file}. The logic scales automatically with the number of processes. The developer uses the \texttt{rank} variable to distinguish between the root (Control) and the others (Targets). The distributed CNOT is abstracted and the \texttt{expose} primitive is used to create the shared state in a single call.
\textit{Complexity:} $O(1)$ in terms of file management. The code remains constant regardless of $N$.

\textbf{2. NetQASM SDK (SquidASM / SimulaQron):}
As these tools operate on a role-based paradigm, the developer must explicitly define the behavior for each node. For a 3-node GHZ state, this requires writing three separate application flows (e.g., \texttt{app\_alice.py}, \texttt{app\_bob.py}, \texttt{app\_charlie.py}).
Furthermore, a Remote CNOT is not a native primitive. The developer must manually implement the "telegate" protocol, which involves generating EPR pairs, performing local measurements, sending classical bits via a separate socket, and applying corrections.
\textit{Complexity:} $O(N)$ in terms of source files and code maintenance. Extending the system from 3 to 4 nodes requires writing a completely new script for the new node and updating the root script to handle the new connection.

\textbf{3. NetSquid / SeQUeNCe:}
These are discrete-event simulators, not programming frameworks. Before implementing the GHZ logic, the user must manually construct the network topology in Python: defining quantum memories, fiber lengths, photon sources, and noise models. The algorithm itself is then defined as a protocol class attached to these physical entities.
\textit{Complexity:} Extremely High. The code is dominated by hardware definitions rather than algorithmic logic.

\subsection{Code Implementation Examples}
\label{subsec:code_examples}

To substantiate the complexity analysis, we present code snippets required to implement the GHZ state generation across the evaluated platforms.

\subsubsection{NetQMPI (SPMD Paradigm)}
Listing~\ref{lst:ghz_netqmpi} shows the full implementation in NetQMPI. The logic is defined in a single block. The \texttt{expose} primitive automatically handles the underlying distributed CNOTs and entanglement distribution to create the shared state defined in Eq.~\ref{eq:final_state_post_expose}. Changing the size from 3 to 50 nodes only requires changing the command line argument \texttt{-n 50}.

\begin{lstlisting}[language=Python, caption={NetQMPI: GHZ Generation (Single File)}, label={lst:ghz_netqmpi}]
def main(app_config=None, rank=0, size=3):
    comm = QMPICommunicator(rank, size, app_config)
    
    # 1. Root creates the base qubit
    qubits = []
    q = Qubit(comm.connection)
    if rank == 0:
        q.H() # Superposition
        qubits.append(q)
        
    # 2. Collective Operation
    # The root exposes the qubit to all ranks (control of remote CNOT)
    comm.expose(qubits, root_rank=0)

    # 3. Perform local operations (if my rank is not root)
    if rank != 0:
        qubits[0].cnot(my_qubit)

    comm.unexpose(root_rank=0)
\end{lstlisting}

\subsubsection{NetQASM SDK (Role-Based)}
In the raw SDK, the logic is strictly separated by roles. Listing~\ref{lst:ghz_netqasm} demonstrates the code required to generate a GHZ state among three nodes (Alice, Bob, and Charlie).

Crucially, note that the developer must explicitly instantiate the connection objects (\texttt{EPRSocket} for quantum links and \texttt{Socket} for classical communication) \textit{before} the main logic, which reveals the scalability bottleneck: adding a fourth node (Dave) forced us to:
\begin{enumerate}
    \item Create a completely new file \texttt{app\_charlie.py}.
    \item Modify \texttt{app\_alice.py} to manually instantiate the new sockets for Charlie (\texttt{sock\_charlie}, \texttt{csock\_charlie}) and duplicate the teleportation logic.
\end{enumerate}

\begin{lstlisting}[language=Python, caption={NetQASM SDK: GHZ Implementation (Alice, Bob, Charlie)}, label={lst:ghz_netqasm}]
# --- app_alice.py (Root) ---
# 1. Explicit Socket Creation
# Alice needs separate sockets for each target
sock_bob = EPRSocket("Bob")
sock_charlie = EPRSocket("Charlie")
csock_bob = Socket("Alice", "Bob")       # Classical
csock_charlie = Socket("Alice", "Charlie") # Classical

with NetQASMConnection("Alice", epr_sockets=[sock_bob, sock_charlie]) as c:
    q = Qubit(c)
    q.H()

    # 2. Manual Remote CNOT to Bob
    epr_b = sock_bob.create_epr()[0]
    q.cnot(epr_b)
    m_b = epr_b.measure()
    c.flush()
    csock_bob.send(str(m_b)) # Send correction bits

    # 3. Manual Remote CNOT to Charlie
    # Logic must be duplicated for the new node
    epr_c = sock_charlie.create_epr()[0]
    q.cnot(epr_c)
    m_c = epr_c.measure()
    c.flush()
    csock_charlie.send(str(m_c))

# --- app_bob.py (Target 1) ---
sock_alice = EPRSocket("Alice")
csock_alice = Socket("Bob", "Alice")

with NetQASMConnection("Bob", epr_sockets=[sock_alice]) as c:
    q = Qubit(c)
    epr = sock_alice.recv_epr()[0]
    epr.cnot(q)
    c.flush()

# --- app_charlie.py (Target 2) ---
# New file required for the third node
sock_alice = EPRSocket("Alice")
csock_alice = Socket("Charlie", "Alice")

with NetQASMConnection("Charlie", epr_sockets=[sock_alice]) as c:
    q = Qubit(c)
    epr = sock_alice.recv_epr()[0]
    epr.cnot(q)
    c.flush()
\end{lstlisting}

\subsubsection{NetSquid (Discrete-Event / Hardware)}
NetSquid operates at a lower level of abstraction. Before implementing the GHZ protocol, the user must define the physical hardware. Listing~\ref{lst:ghz_netsquid} illustrates the verbosity of setting up just the nodes and channels. The actual algorithm requires defining a class inheriting from \texttt{NodeProtocol} for each entity.

Due to the excessive length and technical complexity of the full implementation, the remaining algorithmic code is omitted. It is crucial to emphasize that this comparison does not aim to undermine the utility of discrete-event simulators like NetSquid, but rather to highlight the conceptual gap between physical layer modeling and high-level application development. These tools serve fundamentally different purposes: while discrete-event engines are indispensable for validating hardware fidelity and noise models, they are not designed for algorithmic prototyping. Thus, the level of detail presented here demonstrates that these are not entities directly comparable to NetQMPI, but instead underlying layers that our proposed approach aims to abstract away.

\begin{lstlisting}[language=Python, caption={NetSquid: Hardware Setup}, label={lst:ghz_netsquid}]
# 1. Define Hardware Components
qproc_alice = Node("Alice", num_positions=2,
    mem_noise_models=...)
qproc_bob = Node("Bob", num_positions=2, ...)

# 2. Define Physical Connections
channel_a_b = QuantumChannel("A->B", length=20, 
    models={"delay_model": FibreDelayModel()})
    
# 3. Network Configuration
network = Network("GHZ_Net")
network.add_node(qproc_alice)
network.add_connection(qproc_alice, qproc_bob, 
    channel_to=channel_a_b)

# 4. Protocol Implementation (Not shown: ~100 lines)
# Requires defining state machines for photon emission
\end{lstlisting}

\subsubsection{QuNetSim (High-Level Simulator)}
QuNetSim provides a centralized view similar to a network orchestrator. While the algorithmic part is concise, it requires significant configuration code to set up the simulation environment. The developer must manually instantiate hosts, configure the network topology, establish connections, and manage the start/stop lifecycle of the simulation engine (see Listing~\ref{lst:ghz_qunetsim}). Furthermore, this code runs purely as a simulation and cannot be compiled to real hardware instructions.

\begin{lstlisting}[language=Python, caption={QuNetSim: Network Setup \& Orchestration}, label={lst:ghz_qunetsim}]
# 1. Manual Network Setup
network = Network.get_instance()
alice = Host('Alice')
bob = Host('Bob')
charlie = Host('Charlie')

# 2. Manual Connection Management
alice.add_connection('Bob')
alice.add_connection('Charlie')
bob.add_connection('Charlie') # ... and others ...
alice.start(); bob.start(); charlie.start()
network.start()

# 3. Protocol: Manual Circuit Construction
def ghz_protocol_alice(host):
    # a. Create Root Qubit
    q_root = Qubit(host)
    q_root.H()
    
    # b. Entangle and Send to Bob
    q_bob = Qubit(host)
    q_root.cnot(q_bob) # Local CNOT

    host.send_qubit('Bob', q_bob) 
    
    # c. Entangle and Send to Charlie
    q_charlie = Qubit(host)
    q_root.cnot(q_charlie)
    host.send_qubit('Charlie', q_charlie)
    
    # ... Wait for acknowledgments ...

def ghz_protocol_target(host):
    # Targets must explicitly receive the qubit
    q = host.get_data_qubit('Alice')
    print(f"{host.host_id} received part of GHZ")

# 4. Execution
t1 = alice.run_protocol(ghz_protocol_alice)
t2 = bob.run_protocol(ghz_protocol_target)
t3 = charlie.run_protocol(ghz_protocol_target)
t1.join(); t2.join(); t3.join()
network.stop(stop_hosts=True)
\end{lstlisting}

\subsection{Quantitative Analysis: lines of code (LOC)}

To evaluate the development effort and scalability of the proposed solution, we analyzed the implementation of the GHZ generation protocol across the selected platforms. We utilize \gls*{LOC} as a metric for software complexity, counting only logical lines (excluding comments and imports).

Table~\ref{tab:loc_comparison} summarizes the asymptotic complexity required for both setting up the network topology and implementing the computational logic.

\begin{table}[tbp]
\centering
\setlength{\tabcolsep}{4pt} 
\begin{tabular}{lcccc}
\toprule
\textbf{Framework} & \textbf{Files} & \thead{\textbf{LOC} \\ \textbf{(Network)}} & \thead{\textbf{LOC} \\ \textbf{(Comp.)}} & \textbf{Paradigm} \\
\midrule
\textbf{NetQMPI} & \textbf{1} & $\bm{\mathcal{O}(1)}$ & $\bm{\mathcal{O}(1)}$ & \textbf{SPMD (Abstract)} \\
QuNetSim & 1 & $\mathcal{O}(N^2)$ & $\mathcal{O}(1)$ & Centralized Sim \\
NetQASM SDK & $N$ & $\mathcal{O}(N^2)$ & $\mathcal{O}(N)$ & Role-based \\
NetSquid & $N$ & $\mathcal{O}(N^2)$ & $\mathcal{O}(N)$ & Hardware Descr. \\
\bottomrule
\end{tabular}
\caption{Comparison of Lines of Code (LOC) complexities required for setting up the network topology and implementing the GHZ generation as the number of nodes $N$ increases.}
\label{tab:loc_comparison}
\end{table}

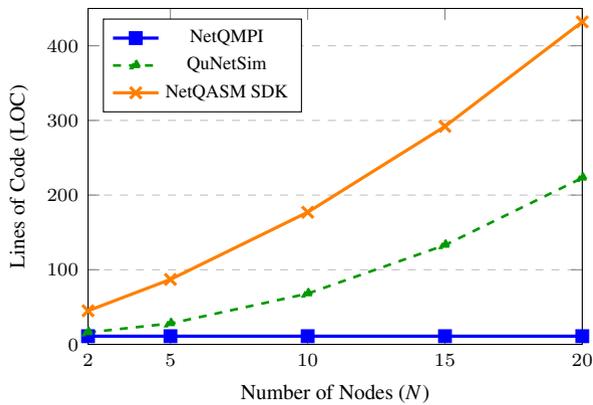
\begin{figure}[tbp]
\centering
\begin{tikzpicture}
\begin{axis}[
    width=0.95\linewidth,
    height=6cm,
    xlabel={Number of Nodes ($N$)},
    ylabel={Lines of Code (LOC)},
    xmin=2, xmax=20,
    ymin=0, ymax=450,
    xtick={2,5,10,15,20},
    ytick={0,100,200,300,400},
    legend pos=north west,
    ymajorgrids=true,
    grid style=dashed,
    legend style={font=\scriptsize},
    label style={font=\footnotesize},
    tick label style={font=\scriptsize},
    cycle list name=color list 
]

\addplot[
    color=blue,
    mark=square*,
    line width=1.2pt
    ]
    coordinates {
    (2,11)(5,11)(10,11)(15,11)(20,11)
    };
    \addlegendentry{NetQMPI}

\addplot[
    color=green!60!black,
    mark=triangle*,
    line width=1.0pt,
    dashed
    ]
    coordinates {
    (2,16)(5,28)(10,68)(15,133)(20,223)
    };
    \addlegendentry{QuNetSim}

\addplot[
    color=orange,
    mark=x,
    line width=1.2pt,
    mark size=3pt
    ]
    coordinates {
    (2,45)(5,87)(10,177)(15,292)(20,432)
    };
    \addlegendentry{NetQASM SDK}

\end{axis}
\end{tikzpicture}
\caption{Comparison of Lines of Code (LOC) required to implement the GHZ state generation protocol as the network size ($N$) increases. NetQMPI maintains constant complexity due to its SPMD collective operations.}
\label{fig:loc_scaling}
\end{figure}

\textbf{Trend Analysis and Scalability:}
The projection of code growth as the network size $N$ increases is illustrated in Figure~\ref{fig:loc_scaling}. The analysis reveals three distinct trends:

\begin{itemize}
    \item \textbf{Linear/Quadratic Growth (State of the Art):} In frameworks like the NetQASM SDK or NetSquid, the complexity grows significantly with $N$. Each new node requires the creation of separate source files, manual instantiation of connection sockets ($\mathcal{O}(N^2)$ for fully connected topologies), and duplication of control logic. This verbose approach forces the developer to write boilerplate code that is essentially identical but structurally necessary, dramatically increasing the probability of introducing copy-paste errors or synchronization bugs.
    
    \item \textbf{Constant Trend (NetQMPI):} In contrast, NetQMPI maintains a flat complexity curve ($\mathcal{O}(1)$). Thanks to the SPMD paradigm and collective primitives, the code written for $N=3$ is identical to the code for $N=100$. The logic is encapsulated in loops (e.g., \texttt{expose}) that iterate over the communicator size dynamically. 
\end{itemize}

\textbf{Implications for Quantum Software Engineering:}
This reduction in code complexity is critical for the development of distributed quantum applications. By decoupling the program logic from the network size, NetQMPI allows for rapid prototyping and ensures a bug-free implementation of complex protocols. In a domain where debugging distributed quantum states is notoriously tricky, minimizing the codebase surface reduces the potential for implementation errors, allowing researchers to focus on algorithmic correctness rather than infrastructure management.
\section{Conclusions}
\label{sec:conclusions}

This work has introduced NetQMPI, a high-level software framework designed to bridge the gap between abstract quantum algorithm design and the low-level management of distributed quantum network resources. By abstracting the verbose primitives of the NetQASM SDK, NetQMPI successfully reduces the complexity of programming distributed quantum applications, offering a scalable and robust development environment.

The primary conclusions extracted from this work are summarized as follows:

\begin{itemize}
    \item \textbf{SPMD paradigm:} NetQMPI facilitates a transition from the labor-intensive, role-based programming model of existing SDKs to a Single Program Multiple Data (SPMD) paradigm. By unifying the control logic into a single source file where roles are defined by the process \texttt{rank}, we have demonstrated that the complexity of setting up a quantum network drops from $\mathcal{O}(N^2)$ to $\mathcal{O}(1)$ in LOC number. This decoupling of program logic from network size is a fundamental requirement for scalable quantum software engineering.

    \item \textbf{Abstraction of the Network Layer:} The library introduces the \texttt{QMPICommunicator} and semantic point-to-point primitives (\texttt{qsend}, \texttt{qrecv}) that completely hide the physical layer's intricacies. The developer is no longer required to manually manage \texttt{EPRSocket} lifecycles, entanglement fidelity, or the synchronization of classical control signals, allowing them to focus entirely on the algorithmic flow of quantum information.

    \item \textbf{Implementation of Collective Operations:} We have provided practical implementations for collective communication in a quantum context. Notably, the introduction of the \texttt{expose} and \texttt{unexpose} primitives addresses the physical constraints imposed by the No-Cloning Theorem. These operations allow for the creation of shared multipartite entangled contexts (such as GHZ states) to simulate broadcasting behavior without violating quantum mechanical principles.

    \item \textbf{Operational Implementation of QMPI:} While previous proposal for a QMPI remain theoretical specifications, NetQMPI provides a functional software implementation. This work demonstrates the practical viability of the MPI paradigm for quantum networks, moving beyond abstract definitions to provide a working execution environment. Crucially, the implementation resolves the conflict between classical broadcasting semantics and the No-Cloning Theorem, proving that high-level distributed abstractions can be effectively mapped to rigorous physical instructions via the NetQASM ecosystem.

    \item \textbf{Integration with the quantum ecosystem:} A critical objective of NetQMPI is its seamless integration with the existing quantum stack. By compiling down to the standardized NetQASM ISA, NetQMPI acts as a backend-agnostic middleware. This allows researchers to execute high-level code on state-of-the-art simulators, such as \textbf{SquidASM} and \textbf{NetSquid}, leveraging their rigorous physical noise models without facing their steep learning curves. Furthermore, this architecture ensures that applications written today in NetQMPI will be executable on future physical quantum hardware that supports the NetQASM standard.
\end{itemize}

In summary, NetQMPI represents a significant step forward in quantum software engineering. It enables developers to write readable, maintainable, and bug-free distributed quantum applications, paving the way for the implementation of complex distributed algorithms.

\printglossary[type=\acronymtype]

\bibliographystyle{IEEEtran}
\bibliography{bib/mendeley}

\begin{IEEEbiography}[{\includegraphics[width=1in,height=1.25in,clip,keepaspectratio]{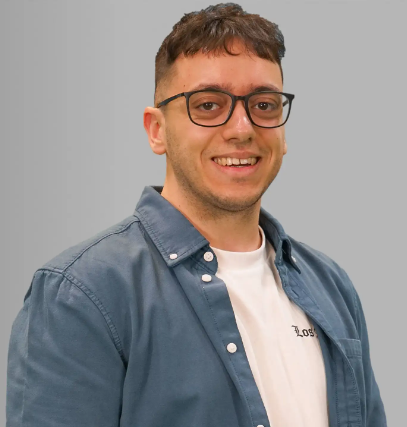}}]{F. Javier Cardama} (Student Member) received the B.S. degree in Computer Engineering from the University of Santiago de Compostela (USC), Spain, in 2021, and the M.S. degree in High-Performance Computing from the same university in 2022. He is currently pursuing the Ph.D. degree in Information Technology Research at the Centro Singular de Investigación en Tecnoloxías Intelixentes (CiTIUS), USC, holding a Xunta de Galicia fellowship. His research interests focus on distributed quantum computing and high-performance computing, specifically on the development of software frameworks and architectures for the integration of quantum technologies in HPC centers.
\end{IEEEbiography}

\begin{IEEEbiography}[{\includegraphics[width=1in,height=1.25in,clip,keepaspectratio]{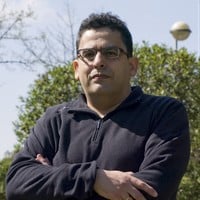}}]{Tomás F. Pena} (Senior Member) received the Ph.D.
in Physics from the University of Santiago de Compostela (USC), Santiago, Spain, in 1994. He is currently Full Professor with the Department of Electronics and Computer Science, USC, and a Senior Member of the Research Center in Intelligent Technologies (CiTIUS), Santiago de Compostela, Spain. His research interests include distributed quantum computing, high-performance computing, parallel systems architecture, optimization of performance in irregular codes and with sparse matrices, and Big Data technologies.
\end{IEEEbiography}

\EOD

\end{document}